# CUMULUS PARAMETERIZATION: THOSE WHO *CAN* REMEMBER THE PAST ARE CONDEMNED TO REPEAT IT

A. D. Del Genio, *NASA Goddard Institute for Space Studies, New York, NY (anthony.d.delgenio@nasa.gov)*

**Introduction:** Moist convection plays a leading role in the dynamics and energy budget of Earth's tropics and influences the sensitivity of Earth's climate to greenhouse gas increases (Zhao, 2014). Because individual convective cells are much smaller than the gridboxes of 3-dimensional global climate models (GCMs), these models parameterize the effects of an ensemble of moist convective updrafts and downdrafts on the environment. Cumulus parameterization has been a focus of the terrestrial meteorology community for half a century. Only in the past decade, however, have GCMs with moist convective physics been applied to other planets (e.g., Mitchell et al., 2011; Leconte et al., 2013; Urata and Toon, 2013). Given our lack of detailed knowledge about convective clouds except on Earth, planetary GCMs are often designed with very simple approaches to cumulus parameterization, adopted from the earliest generations of terrestrial GCMs. These parameterizations were based on breakthroughs in understanding of convection in their time. However, at the same time that planetary GCMs have begun to emerge, a quiet revolution in how we think about terrestrial convection has started to influence the design of terrestrial GCMs. In this paper we review how some of the assumptions in the classical cumulus parameterizations used in planetary GCMs (and still in some terrestrial GCMs) have given way in recent years to a deeper understanding of how convection operates and why it matters for atmospheric dynamics and climate – on any planet.

**Assumptions in traditional parameterizations:** We describe below three of the most common assumptions made in the early history of cumulus parameterization development.

*Undilute "hot towers".* Riehl and Malkus (1958) explained the role of deep convective cells in maintaining Earth's Hadley circulation (in fact, these "hot towers" *are* the upwelling branch of the Hadley cell – air mostly subsides outside active convective updrafts in the tropics). It was long assumed that in order for these cells to reach the tropopause and explain why its moist static energy is similar to that at Earth's surface, buoyant air parcels must rise without interacting with the drier air through which they rise.

*Moist convective adjustment.* Despite this, the first GCM cumulus parameterization (Manabe and Wetherald 1967) ignored the hot tower idea and assumed that convection occurs by mixing between two adjacent conditionally unstable and saturated layers and condenses enough water to restore a moist adiabatic lapse rate and saturation (referred to as "hard" adjustment; a "soft adjustment" version uses 80% relative humidity rather than saturation).

*Quasi-equilibrium.* Arakawa and Schubert (1974) proposed a spectrum of convective cells of different depths (including both undilute and entraining plumes) that responds quickly to destabilization by large-scale processes (vertical motions, radiative heating/cooling). It adjusts the tropical atmosphere to lie along a family of states in the phase space of boundary layer humidity and tropospheric lapse rate that represent a given degree of instability of the column (the more humid the air near the surface, the closer the lapse rate to the moist adiabat). In this sense, the instantaneous environmental thermodynamic structure and resolved-scale forcing determine the instantaneous convective response.

**Recent advances in thinking about convection:** Most of GCM cumulus parameterization history has focused on convective heating and how it adjusts the temperature profile toward neutral stability. The effects of convection on humidity (via detrainment of saturated updraft air as updrafts lose buoyancy, drying by compensating environmental subsidence, and evaporation of rain) were an afterthought.

Derbyshire et al. (2004*)* started a revolution in thinking about convection by comparing the response of cloud-resolving models (CRMs) and single column models (SCMs) that contained GCM parameterized physics to a typical tropical sounding. Relative humidity above the boundary layer in these experiments was varied from humid to dry. The CRMs switched from deep to shallow convection as the troposphere became drier, but the SCMs produced deep convection in all cases. The conclusion was that entrainment of sub-saturated environmental air into the rising updraft, which reduces its buoyancy by evaporating cloud liquid water, was more important than had been previously assumed.

These results invalidate the undilute hot tower idea and show that the atmosphere can depart from quasi-equilibrium, since boundary layer humidity, tropospheric lapse rate and large-scale forcing alone do not determine the convection that results. They are also inconsistent with the concept of moist convective adjustment, which assumes that convection is triggered only in humid air and only occurs in conditionally unstable layers. In reality, convection can be present at all humidities in an unstable atmosphere, varying only in depth; in humid air it pene-trates much deeper than the depth of the conditionally unstable layer (~600 hPa in Earth's tropics).

**Chronic convection biases in Earth GCMs:** The tendency of most cumulus parameterizations to assume an entrainment rate that is too weak causes them to underpredict shallow and midlevel (congestus) convection, overpredict deep convective depths, and overpredict the cumulus mass flux at high altitudes. This creates several types of persistent errors common to most operational GCMs. For example, most GCMs have a warm bias (Kim et al., 2013) and a moist bias (Jiang et al., 2012) near and above the tropical tropopause as a result of detraining too much saturated air and ice and producing too much compensating subsidence warming at these levels.

Likewise, most GCMs tend to trigger deep convection early in the day during the warm season over land, producing peak rain at noon, when insolation is strongest (Dai, 2006; Dirmeyer et al., 2012). Satellite rainfall data show a peak in late afternoon or evening instead. Partly this is an entrainment problem – CRMs indicate that entrainment is strong early in the day, when convection is shallow, and only decreases to weaker values in the afternoon after convection has already begun to deepen (Del Genio and Wu, 2010). In this case, though, another factor comes into play – the tendency for convection to organize into mesoscale clusters when shear and humidity conditions are favorable (Schumacher and Houze, 2006). The organization process perpetuates convection, shifting the rainfall peak to later in the day, and allows convection to propagate.

Another example is the Madden-Julian Oscillation (MJO). The MJO is the primary cause of subseasonal rainfall variability in Earth's tropical warm pool region, but it is absent in most GCMs (Jiang et al., 2015). Many theories have been advanced to explain the MJO, which alone among observed large-scale equatorial wave modes is not predicted by classical shallow water theory on an equatorial β-plane (Matsuno, 1966; Wheeler and Kiladis, 1999). The most promising explanation appears to be that the MJO is an example of a *moisture mode* – an oscillation whose existence is fundamentally tied to prognostic variations in the humidity field in a weak temperature gradient environment and that is driven by sources of moist static energy (e.g., Sobel and Maloney, 2013). Most GCMs apparently fail to produce the MJO because in the unstable but relatively dry conditions ahead of the MJO peak rain period, GCMs prematurely produce deep convection and top-heavy diabatic heating that exports moist static energy, rather than shallow convection that produces bottom-heavy heating, moistens the lower troposphere, and allows surface evaporation and/or large-scale advection to import moist static energy and build instability over several weeks.

For these reasons, convection sensitivity to free tropospheric humidity is now considered to be the most stringent test of cumulus parameterizations available. Unfortunately, there appears to be little awareness of this issue to date in planetary GCM studies, which have either relied on grid-scale resolved moist processes to suffice or have assumed that the simplest cumulus parameterization is the best, without asking what minimum set of requirements should be met by any parameterization.

**Potential planetary implications:** Until planetary GCM studies begin to incorporate analyses of the humidity field, it is impossible to draw any conclusions about whether and how recent terrestrial experience might translate into deeper insights on outstanding issues in planetary science, but we can speculate on several areas worthy of investigation.

*Titan's seasonal cycle:* On Titan, the condensing volatile is methane rather than water. Short-lived cellular clouds reminiscent of convection have been observed at Titan's south pole and in southern midlatitudes during the first half of the Cassini mission, when Titan was in southern summer, and a dramatic outbreak of equatorial organized convection with apparent heavy precipitation that temporarily modified the surface was seen near vernal equinox (Turtle et al., 2011a,b). It has been demonstrated that moist physics is necessary for Titan GCMs to simulate a mean meridional circulation that produces rising motion at the latitudes at which convective clouds are observed (e.g., Mitchell et al., 2011; Schneider et al., 2012). However, Titan GCMs predicted that as Titan shifted into northern spring, analogous outbreaks of convection should have begun in the northern midlatitudes and polar regions. To date, there has been almost no cloud activity observed in Titan's northern hemisphere other than an occasional "lake-effect" cloud in the north polar "lake district."

The only direct measurement of Titan's tropospheric methane profile was at the Huygens probe entry site (Niemann et al., 2005). It shows a well-mixed boundary layer and saturated humidity with respect to a $CH_4$-$N_2$ mixture above (Griffith, 2009). If this is typical of the planet as a whole, then the absence of convection in the northern hemisphere in spring is difficult to explain with any cumulus parameterization, since entrainment has no effect on updraft buoyancy when the air entrained is already saturated, other than a minor direct effect on temperature due to the small cloud-environment temperature difference. However, recent analyses of Cassini radio occultation profiles (Tokano, 2014) and ground-based near-infrared spectra (Ádámkovics et al., 2015) suggest that humidity decreases from the southern to the northern hemisphere. The Mitchell et al. (2011) and Schneider et al. (2012) GCM studies use the Frierson (2007a) cumulus parameterization, which relaxes the atmosphere toward a specified relative humidity and allows for shallow convection to prevent excessive instability. This scheme is an improvement over moist convective adjustment, but it does not produce an MJO (Frierson, 2007b), so it is not clear whether it triggers convection correctly in subsaturated conditions or whether it might produce different Titan seasonality if the reference humidity were varied.

*Saturn's great storms:* The most obvious examples of moist convection on Saturn during the Cassini era have been the dramatic "great storms" that appeared early in the mission in the southern hemisphere and later in the northern hemisphere (Dyudina et al., 2013), both of them in westward jet regions. How often such storms should occur is a major question. Saturn's weak internal heat flux relative to the solar heating of Earth's surface dictates that the convection required to remove heat to upper levels where it can be radiated to space should be rarer on Saturn. But how thermodynamic conditions evolve on Saturn to produce only very sporadic major outbreaks of convective activity remains a mystery. Li and Ingersoll (2015) argue that the greater molecular weight of water relative to that of Saturn's $H_2$-He atmosphere suppresses the development of deep convection.

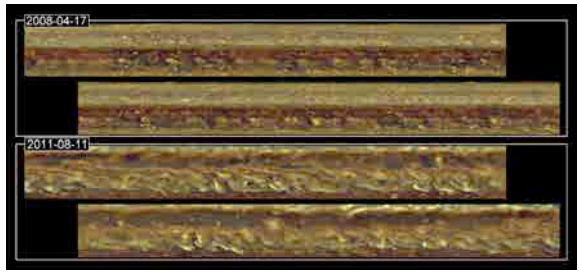

*Fig. 1. Cassini images of a northern hemisphere westward jet region before (2008-04-17) and after (2011-08-11) the development of a giant storm. Picture courtesy JPL.*

Another possibility, though, is that the sensitivity of convection to tropospheric humidity plays the same role on Saturn as it does in Earth's tropics. Analysis of continuum and methane band images of the northern hemisphere westward jet region before the storm outbreak (Fig. 1) indicates that the dominant cloud type in these regions at most times is small cellular clouds that do not reach the upper troposphere, suggesting the presence of shallow convection rising into a dry environment that hinders vertical development (Del Genio and Barbara, 2015). Might the 20-30 year intervals between giant storms at these latitudes simply be a buildup period of moist static energy, analogous to the suppressed period of the MJO on Earth? What kind of cumulus parameterization would a giant planet GCM require to reproduce the sporadic nature of the giant storms?

*Inner edge of the habitable zone*: A basic problem in exoplanet science is the physics of the inner edge of the habitable zone. 1-D models that assume an adiabatic, saturated troposphere (Kopparapu et al., 2013) put the inner edge at T = 340 K, corresponding to an Earth-Sun distance of only d = 0.99 AU, barely closer than Earth's present position. This corresponds to the estimated "moist greenhouse" limit at which the stratospheric $H_2O$ mixing ratio $\sim 3 \times 10^{-3}$, sufficient for photolysis of $H_2O$ and subsequent hydrogen escape to cause all water to be lost over the age of the solar system.

GCMs give a more liberal estimate of the inner edge of the habitable zone than 1-D models but disagree greatly with each other about how surface temperature and stratospheric water increase with increasing insolation (Leconte et al., 2013; Wolf and Toon, 2014, 2015; Yang et al., 2014). Does cumulus parameterization play a role in this disagreement? The most sensitive GCM (Leconte et al., 2013) uses moist convective adjustment, while the others, which are less sensitive, use quasi-equilibrium schemes. As discussed above, the choice of cumulus parameterization can affect the humidity profile and the depth of convection, so it is plausible that convection plays a role in the differing conclusions of these models. None of the models applied to the inner edge problem thus far have a realistic approach to convective entrainment.

**Current trends in cumulus parameterization:** Primarily in response to the absence of an MJO in GCMs, many modeling groups have experimented with increasing entrainment to produce a tighter link between humidity and convection. Stronger entrainment consistently produces an MJO, but often at the expense of a degraded mean state. This can be alleviated, though, by removing radiation imbalances and by strengthening convective downdrafts (Del Genio et al., 2012).

One thing that improves with stronger entrainment is the vertical thermodynamic structure: The tropopause cools and the upper troposphere/lower stratosphere dries (Fig. 2), in better agreement with observations and reanalyses. This can be understood in terms of the reduction in convective depth that results from stronger entrainment, and the stronger tropopause cold trap that results. In the moist greenhouse theory of the inner edge of the habitable zone, erosion of the cold trap with warming is the avenue by which stratospheric $H_2O$ buildup occurs and eventual catastrophic water loss begins. It therefore makes sense to ask whether the choice of cumulus parameterization (including the default saturated adiabat used by 1-D models) has implications for the resulting inner edge estimate.

Although stronger entrainment produces an MJO in terrestrial GCMs, the transition from shallow to deep convection at the onset of the disturbed period of the MJO is also a transition from strong to weak entrainment. The same is true for the afternoon onset of rain in the continental warm season diurnal cycle.

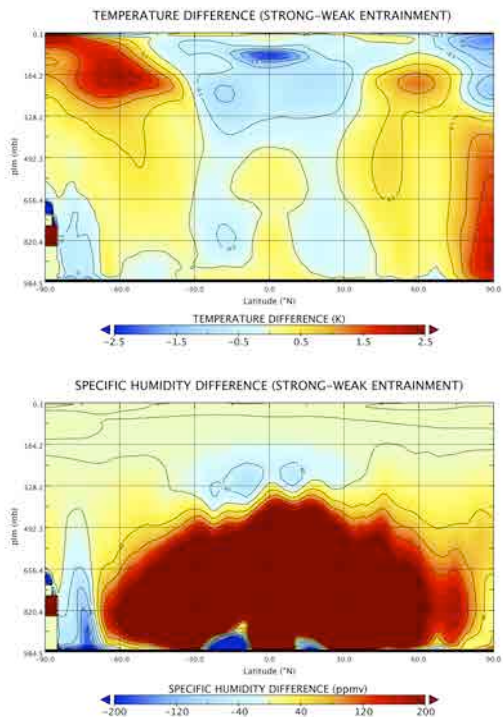

*Fig. 2. Zonal mean (upper) temperature and (lower) specific humidity differences between a version of the GISS GCM with stronger convective entrainment and a baseline version with weaker entrainment.*

CRMs show that as the transition begins, downdrafts bring cold air to the surface, forming *cold pools* that spread, converge with warm humid air, and force uplift in larger eddies that entrain less, leading to further convection. This is another departure from quasi-equilibrium - convection now has "memory" of previous events rather than being determined solely by the current state and forcing. Parameterizing cold pools in GCMs is in its infancy but shows promise as a physically-based way to allow the GCM to entrain strongly much of the time, but more weakly when conditions are favorable for convection to become vigorous and organize on larger scales. A version of the Goddard Institute for Space Studies (GISS) GCM that parameterizes cold pools produces realistic MJO hindcasts and a convection depth vs. column water vapor behavior that better matches data than the standard model (Del Genio et al. 2015). This parameterization is expected to be included in the next generation GISS Earth climate GCM and the planetary version that will be used to simulate past climates of Earth, Venus, and Mars as well as rocky exoplanets.

**Acknowledgements:** This research was supported by the NASA Planetary Atmospheres Program. The results reported herein benefitted from information exchange within NASA's Nexus for Exoplanet System Science (NExSS) research coordination network sponsored by NASA's Science Mission Directorate.